\documentstyle[12pt,epsfig]{article}
\font\tenrm=cmr10

\font\elevenbf=cmbx10 scaled\magstep 1
\font\elevenrm=cmr10 scaled\magstep 1

\renewenvironment{thebibliography}[1]
 { \elevenrm
   \begin{list}{\arabic{enumi}.}
    {\usecounter{enumi}     \setlength{\parsep}{0pt}
     \setlength{\itemsep}{3pt} \settowidth{\labelwidth}{#1.}
     \sloppy
    }}{\end{list}}
\begin{document}
\begin{center}
\vglue 1cm
{\Large  Nuclear Bound States of Antikaons,\\
    or Quantized Multiskyrmions? \\}
\vglue 0.5cm
{Vladimir Kopeliovich$^a$\footnote{{\bf e-mail}: kopelio@inr.ru}  and Irina Potashnikova$^b$\footnote{{\bf e-mail}: irina.potashnikova@usm.cl} \\
a) Institute for Nuclear Research, Russian Academy of Sciences,\\ Moscow 117312, Russia\\
b) Departamento de F\'{\i}sica, Centro de Estudios Subat\'omicos,\\ y Centro
Cient\'ifico - Tecnol\'ogico de Valpara\'iso,\\ Universidad T\'ecnica
Federico Santa Mar\'{\i}a, Casilla 110-V, Valpara\'iso, Chile}

%Universidad T\'ecnica Federico Santa Maria, Casilla 110-V, Valparaiso, Chile}
\end{center}
{\rightskip=2pc
 \leftskip=2pc
\begin{abstract}
{\tenrm\baselineskip=10pt The spectrum of strange multibaryons is considered within the chiral 
soliton model using one of several possible $SU(3)$ quantization models (the bound state rigid oscillator 
version). The states with energy below that of antikaon
and corresponding nucleus can be interpreted as antikaon-nucleus bound states. In the formal limit of small
kaon mass the number of such states becomes large, 
for real value of this mass there are at least several states with positive and negative parity
in the energy gap of one kaon mass. For large values of binding energies interpretation 
of such states just as antikaon-nuclear bound states
becomes more ambiguous.}
\end{abstract}
 \noindent
\vglue 0.2cm}

\baselineskip=14pt
\section{Introduction} 
The studies of multibaryon states with different values of flavor quantum numbers
are of permanent interest. They are closely related to the problem of existence of strange 
quark matter and its fragments, strange stars (analogs of neutron stars).
Besides traditional approaches to this problem based usually on the potential and/or 
quark models, the chiral $SU(3)$ dynamics, mean field theories, etc., the chiral soliton approach (CSA) proposed by Skyrme \cite{skyrme}
is effective and has certain advantages before conventional methods (some early 
descriptions of this model can be found in \cite{hosc}). 
The quantization of the model performed first in the $SU(2)$ 
configuration space for the baryon number one states \cite{anw}, somewhat later for configurations with
axial symmetry \cite{vkax, bc} and for multiskyrmions \cite{c3,w4,irw}, allowed, in particular, to describe
the properties of nucleons and $\Delta$-isobar \cite{anw} and, more recently,
some properties of light nuclei, including so called "symmetry energy" 
\cite{ksm}\footnote{Recently the neutron rich isotope 
$^{18}B$ has been found to be unstable relative to the decay
$^{18}B \to ^{17}B +n$ \cite{spyrou}, in agreement with prediction of
the CSA \cite{ksm}. This can be considered as an illustration
of the fact that the CSA provides quite realistic 
predictions for the case of nonstrange nuclei.} 
and many other properties \cite{mmw}.

The $SU(3)$ quantization of the model has been performed within the rigid \cite{gua} or soft \cite{sweig} 
rotator approach and also
within the bound state model \cite{cakl}.
The binding energies of the ground states of light hypernuclei have been described
within a version of the bound state chiral soliton model \cite{wk}, in qualitative, even
semiquantitative agreement with empirical data \cite{vkh}.
 The collective motion contributions
have been taken into account here (single particles excitations should be added), and special subtraction 
scheme has been used to remove uncertainties in absolute values of masses intrinsic to the CSA.
It makes sense therefore to extend such investigation to the higher in energy (excited) states, some of them
may be interpreted as antikaon-nuclei bound states.

The antikaon-nuclei interactions and possible bound states of antikaons and nuclei 
have attracted recently much attention
\cite{akayam} - \cite{haimei}. Theoretically deeply bound states of antikaons in nuclei 
have been obtained as a solution of 
many-body problem by Akaishi and Yamazaki \cite{akayam}-\cite{yamak}.
Most recent reviews of this topic within the framework of conventional approaches can be found
in \cite{gal,weise}.
Here we investigate the possibility of interpretation of such states as quantized
multiskyrmions (configuration with baryon number one is called usually a skyrmion).
The spectrum of quantized multiskyrmions is very rich, and some of these 
states are appropriate for interpretation as bound antikaon-nuclei states.

Within the CSA there is a simple argument that at small 
value of the kaon mass $m_K$ there should be quantized states of multiskyrmions with the mass below the sum of masses
of the kaon and corresponding number of nucleons. Indeed, the strangeness (flavor)
 excitation energies are proportional to $m_K^2$, both in the rotator \cite{gua,sweig} and in
the bound state models of skyrmion quantization \cite{cakl}. Therefore,
the mass of any state with baryon number $B$, strangeness $S$, isospin $I$, spin $J$ 
can be presented as sum of two terms
$$ M(B,S,I,J...) \simeq M(B, S=0, ..) + m_K^2 \Gamma_B \;C(B,S,I,J...), \eqno (1) $$
where $\Gamma_B $ is the $\Sigma$-term (see Table 1), $C(B,S,I,J...)$ is some quantity of the order $\sim 1$,
depending on quantum numbers of the system.
Evidently, at small enough $m_K$ the contribution given by $(1)$ is smaller than the sum $M(B, S=0, ..) + |S|m_K$, and the 
number of states with the mass given by $(1)$ in the gap between $M(B, S=0, ..)$ and $M(B, S=0, ..) + |S| m_K$
becomes large. This argument is quite 
rigorous, however, for
realistic value of $m_K$ it is a question of numerical calculation to find out, which
states have the energy below that of the multibaryon 
plus antikaon system (here we consider the case of strangeness $S=-1$).

The interpretation of these states with fixed external quantum numbers in terms of hadronic 
constituents is not straightforward and not unique. Each state is the whole Fock column 
of hadronic components with different weights. We could only, in some particular situations,
make statements about dominance of some components of this Fock column. 

It should be specially pointed out that here we are using one of possible $SU(3)$ quantization  models,
the rigid oscillator version of the bound state model \cite{wk} which seems to be the simplest one.
This quantization scheme can provide quantized states with definite restrictions on allowed
quantum numbers of the states, including their spatial parity. E.g., only positive parity 
baryons appear when the 
basic baryon number $1$ hedgehog-type $SU(2)$ configuration is quantized in this way. 
To get the states 
with negative parity, for example, the low mass $\Lambda(1405)$ state, an actual 
candidate to be the
antikaon-nucleon bound state, one should provide at least second order expansion in 
mesonic fluctuations around
the basic classical configuration (hedgehog). Considerable success in describing the properties
of $\Lambda(1405)$ has been reached in this way in \cite{scscg}.

For the case of multiskyrmions similar approach is technically very complicated and is 
not performed except
few attempts \cite{ssg,scsc}. In the $SU(2)$ case the qualitative description of some dibaryon states was
obtained in \cite{scsc}. Therefore, the expected spectrum of negative strangeness 
states may be considerably
richer than obtained in present paper.

In the next section isotopical properties of the $\bar KNN$ and $\bar KNNN$ systems are briefly discussed.
Section 3 contains description of starting positions of the CSA, in section 4 we recollect the
spectrum of $SU(2)$ quantized dibaryons, section 5 contains the formulas summarizing 
the CSA results for strange (flavored) multiskyrmions, our main results for the spectrum of
strange baryonic states with $B=2$ and $3$ are presented in sections 6,7. Our former 
results for strange dibaryons are recollected in section 6. Excitations of the ground states
of the $B=2$ and $3$ systems in some cases could be interpreted as antikaon-nuclei bound states.
\section{Phenomenology}
The $K^-pp$ cluster has been proposed in \cite{yamak} as a fundamental unit which plays
an important role in formation of similar strangeness $S=-1$ clusters in heavier nuclei.

Here we discuss first some consequences of isotopic invariance of strong interactions
involving strange particles. The state $K^-pp$ which has the 3-d component of isospin
$I_3=1/2$, is in fact a coherent combination of states with isospins $I=3/2$ and $I=1/2$:
$$|K^-pp\rangle = \sqrt{{1\over 3}} |\bar KNN;\,3/2,+1/2\rangle +\sqrt{{2\over 3}}
|\bar KNN;\,1/2,+1/2\rangle. \eqno (2)$$
Another physical state with same quantum numbers is
$$|\bar K^0(pn)_{I=1}\rangle = \sqrt{{2\over 3}} |\bar KNN;\,3/2,+1/2\rangle -\sqrt{{1\over 3}}
|\bar KNN;\,1/2,+1/2\rangle, \eqno (3)$$
where $(pn)_{I=1}$ system has isospin $I=1$. So, same cluster which can be seen in $K^-pp$ system 
should be 
seen also in $\bar K^0(pn)$ system, but with about 4 times smaller probability \footnote{We take
into account that the $pn$ system has isospin $I=1$ with probability $1/2$.}.
The $\bar KNN$ state with isospin $I=3/2$ includes the state with charge $+2$, it is $\bar K^0pp$,
and state with charge $-1$, it is $ K^-nn$.

Another possibility to have the state with isospin $I=1/2$ is to combine antikaon state with the
isospin zero $2N$ state:
$$|\bar KNN;\,1/2,+1/2\rangle = |\bar K\rangle |(pn)_{I=0}\rangle. \eqno (4)$$
In total, we have for the $\bar KNN$ system 8 different components which can be splitted into
quartet (isospin $I=3/2$) and two doublets.
Within the CSA we shall obtain the states with baryon number $2$ and quantum numbers ---
strangeness, isospin, spin --- as indicated above, and estimate their masses. 

Similar for the $B=3$ systems. 
In the case of $\bar KNNN$ system we have in total 16 components which can be 
separated into one quintet with maximal isospin $I=2$, three triplets with $I=1$ and two singlets.
The maximal value of the 3-d component of isospin is $I_3=+2$ ($\bar K^0 ppp$-system), and minimal value is
$I_3=-2$ ($K^-nnn$ system). As it was shown 
previously and we shall see here, within the CSA there is specific dependence of the mass of baryonic system 
on its isospin, usually states with lower isospin have smaller energy.
\section{Basic ingredients and features of the CSA}
 The CSA is based on few 
principles and ingredients incorporated in the {\it truncated}
 effective chiral lagrangian \cite{skyrme,hosc,anw}:
$$L^{eff} = -{F_\pi^2\over 16}Tr l_\mu l_\mu + {1\over 32e^2}Tr [l_\mu l_\nu]^2
+{F_\pi^2m_\pi^2\over 8}Tr \big(U+U^\dagger -2\big)+..., \eqno (5)$$
the chiral derivative $l_\mu = \partial_\mu U U^\dagger,$ $U\in SU(2)$ or $U\in SU(3)$-
unitary matrix depending on chiral fields, 
$m_\pi$ is the pion mass, $F_\pi$- the pion decay constant known experimentally, $e$ - the only 
parameter of the model in its minimal variant
proposed by Skyrme \cite{skyrme}.

The mass term $\sim F_\pi^2m_\pi^2$,
 changes asymptotics of the profile $f$ and the structure
of multiskyrmions at large $B$, in comparison with the massless case. For the $SU(2)$ case
$$ U= cos f + i\, (\vec{n}\vec{\tau}) sin\,f, \eqno (6)$$
the unit vector $\vec{n}$ depends on 2 functions, $\alpha,\;\beta$.
Three profiles $\{f,\;\alpha,\;\beta\}(x,y,z)$ parametrize the 4-component
unit vector on the 3-sphere $S^3$.

The topological soliton (skyrmion) is configuration of chiral fields, 
possessing topological charge identified with the
baryon number $B$ \cite{skyrme}:
$$ B= {1\over 2\pi^2}\int s_f^2 s_\alpha I\left[(f,\alpha,\beta)/(x,y,z)\right] d^3r, \eqno (7)$$\\
where $I$ is the Jacobian of the coordinates transformation, $s_f=sin\,f$. 
So, the quantity $B$ shows how many times the unit sphere $S^3$ is covered when 
integration over 3-dimentional space $R^3$ is made.

The chiral and flavor symmetry breaking term in the lagrangian density depends on kaon mass and decay constant
$m_K$ and $F_K$ ($F_K/F_\pi \simeq 1.23$ from experimental data):
$$ L^{FSB} = {F_K^2m_K^2-F_\pi^2m_\pi^2\over 24}Tr \bigl(U+U^\dagger -2\bigr) \bigl(1-\sqrt 3 \lambda_8\bigr)-$$ 
$$-{F_K^2-F_\pi^2\over 48}Tr \bigl(Ul_\mu l_\mu+ l_\mu l_\mu U^\dagger\bigr) \bigl(1-\sqrt 3 \lambda_8\bigr) 
\eqno(8). $$
This term defines the mass splittings between strange and nonstrange baryons (multibaryons), modifies some
properties of skyrmions and is crucially important in our consideration.

As we have stressed previously, the great advantage of the CSA is that multibaryon 
states --- nuclei, hypernuclei ... --- can be considered on equal footing with the $B=1$ case.
Masses, binding energies of classical configurations, the moments of 
inertia $\Theta_I,\;\Theta_J$..., the $\Sigma$-term (we call it $\Gamma$) and some other
characteristics of chiral solitons contain implicitly  
information about interaction between baryons.
Minimization of the mass functional $M_{class}$ provides 3 profiles $\{f,\alpha,\beta\}(x,y,z)$ and
allows to calculate moments of inertia, etc . 
\section{Mass formula for multibaryons quantized in $SU(2)$}
In the $SU(2)$ case, the rigid 
rotator model (RRM) used at first in \cite{anw} for the $B=1$ case, is most effective and successfull.
It allowed to describe successfully the
properties of nucleons, $\Delta (1232)$- isobar, as well as many properties of light nuclei \cite{mmw}, and also 
mass splittings of nuclear isotopes, including neutron rich nuclides with atomic numbers up to 
$\sim 30$ \cite{ksm}.

When the basic classical configuration possesses definite symmetry properties, the 
interference between iso- and usual space rotations becomes important. We consider here first
an example of the axially symmetrical configuration which is believed to provide the absolute 
minimum
of the classical static energy (mass) for the $B=2$ case.
The mass formula for the axially symmetric configuration has been obtained first for the nonstrange
states in \cite{vkax} and, in greater detail somewhat later, in \cite{bc}:
$$M(B,I,J,\kappa) = M_{cl}+ {I(I+1)\over 2\Theta_I} + {J(J+1)\over 2\Theta_J} +{\kappa^2\over 2}
\left({1\over \Theta_{I,3}}-  {1\over \Theta_{I}}- {4\over \Theta_{J}}\right), \eqno (9)$$
where $\kappa = I_3^{bf}$, $I^{bf}$ and $J^{bf}$ are body-fixed isospin and spin of the
system, and the relation takes place $J_3^{bf} = -2I_3^{bf}$ as a consequence of the
generalized axial symmetry of the $B=2$ classical configuration (see Eq. $(10)$).

This formula is in agreement with known quantum mechanical formulas for the energy of axially
symmetrical rotator \cite{ll}.
The classical characteristics of the lowest baryon numbers states --- moments of inertia $\Theta_I, \Theta_J,
\Theta_{I,3}$, which enter formula $(9)$, as well as other quantities, necessary for calculating the spectrum of $SU(3)$ quantized 
states, are given in Table 1.

\begin{center}
\begin{tabular}{|l|l|l|l|l|l|l|l|}
\hline
$B$& $ \Theta_I$& $\Theta_J$&$\Theta_3$& $\Theta_S$ &$\Gamma $ &$ \omega_S$ &$\mu_S$ \\
\hline
$1 $& $5.55$    &$5.55 $    &$5.55$    & $ 2.04 $   &   $ 4.80$&$306$ &$3.165$\\
\hline
$2 $&$11.47 $   &$19.74$    &$7.38 $   & $4.18$     &   $9.35$ &$293$&$3.081$  \\
\hline
$3 $&$14.4 $    &$49.0$     &$14.4 $   & $6.34$     &   $14.0$ &$289$&$3.066$  \\
\hline
$4 $&$16.8 $    &$78.0$     &$20.3 $   & $8.27$     &   $18.0$ &$283$&$2.972$  \\
\hline
\end{tabular}
\end{center}
{\tenrm {\bf Table 1.}  Characteristics of classical skyrmion configurations
which enter the mass formulas for multibaryons. The numbers are taken from \cite{vkwz,vk01}:
 moments of inertia $\Theta$ and $\Sigma$-term $\Gamma$ -
in units $GeV^{-1}$, $\omega_S$ - in $MeV$, $\mu_S$ is dimensionless (see next sections for
explanation)\footnote{In some
formulas we add lower index $B$ for all quantities to emphasize dependence on baryon number, e.g. 
$\mu_{S,B}$.}. Parameters of the 
model $F_\pi =186\,MeV; e=4.12$ \cite{vkh,vkwz,vk01}. }\\

The rational map approximation \cite{hms} simplifies considerably calculations of various
characteristics of classical multiskyrmions presented in Table 1\footnote{Explicit expressions for the quantities
shown in this table can be found in \cite{vk01} and \cite{mmw}.}. The value of $\Theta_J$ in Table 1  for the baryon numbers $B=3$ and $4$
is the average one of the diagonal elements of the orbital inertia tensor.

Here in  Tables 2,3 we present for completeness the result of the calculation of dibaryons spectrum according to the above formula.
Many of these results have been obtained previously in \cite{bc}; unlike \cite{bc} we pretend to calculate
the differences of masses of states with different quantum numbers, not the absolute values of masses
which are controlled by the loop corrections and/or so called Casimir energy which has been calculated
approximately for the $B=1$ case, see \cite{mous,mewa}. For our choice of the model parameters
the mass differences presented in Tables 2,3 are somewhat smaller (by few tens of MeV) than 
the mass differences which can be extracted from results of \cite{bc} obtained with 
parameters of the paper \cite{anw}.
\begin{center}
\begin{tabular}{|l|l|l|l|l|l|l|}
\hline
$\, I$& $ J$ &$\, \kappa $ & \quad Content& $ \qquad \qquad \Delta E $& $\Delta E \, (MeV)$\\
\hline
$\, 1 $&$ 0$ &$\,0$&$NN (^1S_0) $&$\quad 0 $&$ $  \\
\hline
$\, 0 $&$ 1$ &$\, 0 $&$NN (^3S_1) $& $1/\Theta_J -1/\Theta_I $&$-36  $  \\
\hline
$\,^* 1 $&$ 1$ &$\, 0$&$NN\pi ?$&$1/\Theta_J  $ &$\;\;51 $  \\
\hline
$\, 1 $&$ 2$ &$\, 0$&$NN (^1D_2);\Delta N(^5S_2)$&$3/\Theta_J $&$\;\;153 $ \\
\hline
$\, 0 $&$ 3$ &$\, 0$&$NN(^3D_3);\Delta\Delta(^7S_3)$ &$6/\Theta_J -1/\Theta_I $&$\;\;219$ \\
\hline
$\, 2 $&$ 1$ &$\, 0$&$\Delta N(^3S_1);\Delta N(^3D_1)$ &$1/\Theta_J +2/\Theta_I $&$\;\;225$ \\
\hline
$\, 3 $&$ 0$ &$\, 0$&$\Delta\Delta(^1S_0); NN\pi\pi$ &$5/\Theta_I $&$\;\;435 $ \\
\hline
$\, 2 $&$ 4$ &$\, 2$&$\Delta N(^5D_4);NN\pi$ &$2/\Theta_J +1/\Theta_I+2/ \Theta_{I,3} $&$\;\;462 $\\
\hline
\end{tabular}
\end{center}
{\tenrm {\bf Table 2.} The quantum numbers, possible hadronic content and the energy (in $MeV$) of positive 
parity states above the singlet $NN$ scattering state with $I=1,\;J=0 $.} \\

As it was shown first in \cite{bc}, the parity of such states is
 $$ P_{ax, B=2} = (-1)^\kappa.  \eqno(10)  $$
The deuteron-like state $(I=0,\;J=1)$ has energy by $36\,MeV$ lower than the $NN$ scattering state 
$ I=1,\;J=0)$ \cite{vkax,bc}\footnote{The measured value of deuteron binding energy is $\epsilon_d
\simeq 2.2\,Mev$. Within the CSA the deuteron-like state is lower in energy than the singlet $NN$
scattering state because the orbital inertia $\Theta_J$ is considerably, by a factor $1.5$, greater
than the isotopic moment of inertia $\Theta_I$, see Table 1. This remarkable property takes place
in all known variants of the CSA.}, so, the value $\sim 40\,MeV$ can be considered as uncertainty of our 
predictions of masses in the $SU(2)$ case.

The coefficient after $\kappa^2$ in eq. $(9)$ is negative, therefore, states with maximal possible value
of $|\kappa|$ at fixed $I,\,J$ have the lowest energy, linearly dependent on $I$ and $J$ after cancellation of
quadratic terms:
$$E_{kin} = {I\over 2\Theta_I} + {J\over 2\Theta_J} +{I\over 2\Theta_{I,3}}. \eqno (11) $$
This formula is valid for negative parity states with $I=1,\,J=2, \kappa=\pm 1$, or,
generally, $J=2I,\; \kappa=\pm I$.

\begin{center}
\begin{tabular}{|l|l|l|l|l|l|l|}
\hline
$\, I$& $ J$ &$\, \kappa$ &\quad Content&\qquad \qquad $ \Delta E $& $\Delta E \, (MeV)$   \\
\hline
$\, 1 $&$ 2$ &$\, \pm 1$&$NN (^3P_2) $&$1/\Theta_J -1/2\Theta_I +1/2\Theta_{I,3} $&$\;\;75 $  \\
\hline
$\, 1 $&$ 3$ &$\, \pm 1$&$NN (^3P_3, ^3F_3) $& $4/\Theta_J -1/2\Theta_I +1/2\Theta_{I,3} $&$\;\;229 $  \\
\hline
$\, 2 $&$ 2$ &$\, \pm 1$&$\Delta N (^3P_2);NN\pi$&$1/\Theta_J +3/2\Theta_I +1/2\Theta_{I,3} $&$\;\;249 $ \\
\hline
$\, 2 $&$ 3$ &$\, \pm 1$&$\Delta N (^5P_3);NN\pi$&$4/\Theta_J +3/2\Theta_I +1/2\Theta_{I,3} $&$\;\;402 $ \\
\hline
$\, 2 $&$ 4$ &$\, \pm 1$&$\Delta N (^3F_4);NN\pi$ &$8/\Theta_J +3/2\Theta_I +1/2\Theta_{I,3} $&$\;\;606 $ \\
\hline
\end{tabular}
\end{center}

{\tenrm {\bf Table 3.} The quantum numbers, possible hadronic content and the energy of negative 
parity states above the $NN$ scattering state with $ I=1,\;J=0$.} \\

Some comment is necessary concerning the state with $I=J=1,\;\kappa=0$ which is forbidden by Finkelstein - 
Rubinstein (FR) constraint and cannot decay into the $NN$-pair due to the Pauli principle. Such a state, if 
it exists, is an example of elementary particle with $B=2$, different from ordinary deuteron or 
singlet scattering state consisting
mainly of two nucleons \cite{vknar}. 
Such states have been considered earlier in \cite{mads,masw} where their masses were found to be
higher, greater than $2120\, MeV$.
Experimental situation with possible observation of such states has been described in \cite{gerkhr}.

The energy of such state, shown in Table 2, does not include the possible
difference of Casimir energies (or loop corrections) between FR allowed and FR forbidden states.
If this energy is large, this state, as well as the $I=J=0$ state, should have energy larger than 
shown in Table 2.

Generally, for multiskyrmions the internal constituents --- nucleons, first of all --- are not
identifiable immediately. Some guess and analysis of quantum numbers are necessary for this purpose.
A possible hadronic content of dibaryon states is shown in Tables 2, 3. Evidently, states with the 
value of isospin $I\geq 2$ cannot be made of 2 nucleons only, additional pions are needed, or $\Delta$ 
instead of some nucleons. By same reason the states with $I\geq 2$ cannot be observed in
nucleon-nucleon interactions. The states with isospin $0$ or $1$ could appear as some enhancements in
corresponding partial wave of the $NN$ scattering amplitude.

For configuration with baryon number $B=3$ the symmetry properties of the classical 
configuration important for quantization have been established first by L.Carson \cite{c3}
and exploited recently in \cite{mmw}.
As a consequence of the symmetry properties of classical $B=3$ configuration which has
characteristic tetrahedral shape, the equality between body fixed spin and isospin takes
place, $K=L$. 
The parity of quantized states equals \cite{c3}
$$P=(-1)^{(K_3+L_3)/2} =(-1)^{M_3/2}. \eqno (12)$$
The analysis and interpretation of the $B=3$ states is more complicated than the $B=2$ states,
and only few of them were considered in \cite{c3,mmw}.

%Quantum states with isospin $I\geq 5/2$ cannot be made of 3 nucleons only
%Rational map approximation \cite{hms} allows to establish more clear the properties of skyrmion
%configurations relative to the parity transformation. 
%The action of the parity operator
%in this case is
%$$ P\,z = -{1\over \bar z},\qquad P\,R(z) = -{1\over \bar R(z)}, \eqno (12)$$
%In particular, when $R(z)= z^n$ which corresponds to $B=n$, then we have an equality
%$R(1/\bar z) = 1/\bar R(z)$
\section{Spectrum of multibaryons with strangeness in the rigid oscillator model}  
The observed spectrum of states is obtained by means of quantization procedure
and depends on the quantum numbers and characteristics of skyrmions presented in 
Table 1.

Within the  bound state model (BSM)
\cite{cakl,wk,vkh} antikaon is bound by the $SU(2)$ skyrmion. The mass formula takes place
$$ M = M_{cl} + \omega_S + \omega_{\bar S} + |S| \omega_S + \Delta M_{HFS} \eqno (13)$$
where flavor and antiflavor excitation energies 
$$\omega_S= N_c(\mu_S-1)/8\Theta_S,\;\;\omega_{\bar S}= N_c(\mu_S+1)/8\Theta_S,\eqno (14)$$
$$\mu_S = \sqrt{1+\bar m_K^2/M_0^2}\simeq 1+{\bar m_K^2\over 2M_0^2},$$ 
$$ M_0^2=N_c^2/(16\Gamma\Theta_S)\sim N_c^0,\quad \mu_S \sim N_c^0,\eqno (15)$$ 
$N_c$ is the number of colors of underlying QCD.

The hyperfine splitting correction depending on hyperfine splitting
constants $c_S$, $\bar c_S$, isospin, "strange isospin" $I_S$ and angular momentum $J$ equals in the case
when interference between usual space and isospace rotations is negligible or not important,
is
$$\Delta M_{HFS} = {J(J+1)\over 2\Theta_J} 
+\frac{c_SI_r(I_r+1)- (c_S-1)I(I+1) +(\bar c_S-c_S)
I_S(I_S+1)}{2\Theta_I} \eqno (16)$$

The hyperfine splitting constants are equal
$$c_S=1- {\Theta_I\over 2\Theta_S \mu_S}(\mu_S-1) ,\qquad 
\bar c_S =1 - {\Theta_I\over \Theta_S\mu_S^2}(\mu_S-1), \eqno (17)$$
Strange isospin equals $I_S=1/2$ for $S=\pm 1$, for negative strangeness in most cases
of interest $I_S=|S|/2$ which minimizes this correction (but generally it can be not so).
We recall that body-fixed isospin $\vec I^{bf} = \vec I_r +\vec I_S$, $\vec I_r$ is the isospin
of skyrmion without added antikaons. It is quite analogous to the so called "right" isospin
within the rotator quantization scheme.
When $I_S=0$, i.e. for nonstrange states, $I=I_r$ and this formula goes over into $SU(2)$
formula for multiskyrmions
$$E_{kin} = {J(J+1)\over 2 \Theta_J}  +  {I(I+1)\over 2\Theta_I}, \eqno (18)$$
where we neglect the interference terms \cite{vk01,mmw}.
Correction $\Delta M_{HFS}\sim 1/N_c$ is small at large $N_c$, and also for heavy 
flavors \cite{cakl,vk01}.

For the case of classical state with generalized axial symmetry an additional term appears
$$ \Delta E^{axial} = {\kappa^2\over 2}\left[{1\over \Theta_3} - {1\over \Theta_I} 
-{4\over \Theta_3}\right] = \kappa^2 \delta(\Theta), \eqno (19)$$
which differs for states with different parities (different $\kappa$, see Tables 1,2).

The mass splitting within $SU(3)$ multiplets is important and convenient for us here since the unknown
for the $B\geq 1$ solitons Casimir energy cancels in the mass splittings.
For the difference of energies of states with strangeness $S$ and with $S=0$ 
which belong to the same multiplet $(p,q)$ we obtain using the above expressions for the
constants $c_S$ and $\bar c_S$
$$\Delta E (p,q; I,J,S; I_r,J_0,0) = |S|\omega_S + {\mu_{S,B}-1 \over 4\mu_{S,B} \Theta_{S,B}}
[I(I+1)-I_r(I_r+1)]+$$
$$ + {(\mu_{S,B}-1)( \mu_{S,B}-2) \over 4\mu_{S,B}^2 \Theta_{S,B}}I_S(I_S+1)
+{1\over 2\Theta_J}[J(J+1)-J_0(J_0+1)]+(\kappa^2-\kappa_0^2)\delta(\Theta), \eqno (20)$$
if the underlying classical configuration possesses axial symmetry.
For arbitrary strangeness $I_S\leq |S|/2$, and $J_0=J$ if these states belong to
the same $SU(3)$ multiplet.
The values of the quantities which enter above formulas are shown in Table 1.
\section{Dibaryons with strangeness}
Strange dibaryons have attracted much attention beginning with pioneer papers \cite{jaffe,
mads,masw,goldm,oka}.
Recent discussion of this topic and many important references can be found in \cite{gal}. 
Here we do not discuss
the $S=-2$ H-dibaryon \cite{jaffe} which is the $SU(3)$ singlet and appears as the $SO(3)$ 
soliton within the chiral soliton approach \cite{bal,jakorp}.

For completeness we present here the former results by B.Schwesinger et al \cite{kss2}
for energies of different strange dibaryons within the soft rotator model with 
$SU(3)$ configuration mixing.
\begin{center}
\begin{tabular}{|l|l|l|l|l|l|l|l|l|l|}
\hline
$multiplet $& $\;\;\{\overline{10}\}$&$\;\;\{27\}$ &$\;\{\overline{10}\}$&$\;\;\{27\}$ &$\;\;\{27\}$&$\;\;\{27\}$&$\;\{27\}$&$\;\;\{35\}$ &$\{28\}$  \\
\hline
$\;\;\; S,I$& $-1,\,1/2 $& $-1,\,1/2 $&$-2, \, 1$& $-2, \, 0$&$-3, \,1/2$&$-3, \,3/2$& $-4,\, 0$ &$-5,\,1/2 $ &$-6,\,0$ \\
\hline
\hline
$\;\; state $& \quad $\Lambda N$&\quad $\Lambda N$&$\;\;\Xi N$ &$\;\;\Lambda \Lambda $&$\;\;\Lambda \Xi$&$\;\;\;\Sigma \Xi$ &$\;\Xi\Xi $& $\;\;\;\Xi\Omega$&$\;\Omega\Omega$\\
\hline
$\;\,\Delta E^{SRM}$& \quad $ 30 $  &\quad $ 70 $  &$\;\;100$ &$\;\;110$  &$\;\;140 $ &$\;\;\,\;90 $ & $\;150$  & $\;\;\,40$ &$\;\;30$\\
\hline
\end{tabular}
\end{center}
{\tenrm \bf Table 4.} {\tenrm The energy above threshold $\Delta E$ in $MeV$ for dibaryons 
with $J^P=0^+$, different values of strangeness $S$ and isospin $I$. The $SU(3)$ multiplet, 
which the main component of the dibaryon configuration belongs to, is indicated in the 
upper line.
Calculations made according to the soft rotator model \cite{kss2}}. \\

As can be seen from Table 4, we did not predict in \cite{kss2} the bound states of dibaryons, 
all states of lowest energy shown in this Table are above the corresponding 
two-baryon thresholds (for consistency we took theoretical values of baryon masses which
do not coincide with empirical values).
These lowest states can be and should be interpreted as virtual states, or scattering 
states similar to the $(NN)$ $^1S_0$ scattering state, the so called singlet deuteron.
Presence of such states leads to the enhancement of scattering cross section in
corresponding channel, as seen in the $\Lambda N$ or $\Lambda\Lambda$ data,
(see, e.g. \cite{gal}). In view of considerable numerical uncertainty of these results
there remains still a chance that nearest to threshold dibaryons can be bound.

{\baselineskip=13pt In previous publication on this subject \cite{kss1} we obtained bound dibaryons,
but the poorly known Casimir energies of the order of $N_c^0$ \cite{mous,mewa} (discussed already
in this paper in connection with nonstrange dibaryons)
 have not been taken into account in \cite{kss1}.
In fact, we should write for baryons
$$M_1(p,q;Y,I)= M_1^{class} +\Delta M_1(p,q;Y,I) + M_1^{Cas}  \eqno (21)$$
and for dibaryons (multibaryons in general case)
$$M_2(p,q;Y,I)= M_2^{class} +\Delta M_2(p,q;Y,I) + M_2^{Cas}.  \eqno (22)$$
$\Delta M(p,q;Y,I)$ is the quantum numbers dependent quantum correction, 
hypercharge $Y=B+S$, $M^{Cas}\sim N_c^0$ 
is the Casimir energy or loop correction.
When we calculated the energy (mass) difference
$$\Delta M = M_1(p_1,q_1;Y_1,I_1)+M_1(p_2,q_2;Y_2,I_2)-M_2(p,q;Y,I)=$$
$$=2M_1^{class}-M_2^{class}+\Delta M_1(p_1,q_1;Y_1,I_1)+\Delta M_1(p_2,q_2;Y_2,I_2)-\Delta M(p,q;Y,I)+$$
$$+2M_1^{Cas}-M_2^{Cas}, \eqno (23)$$
in \cite{kss1} we ignored the term $2M_1^{Cas}-M_2^{Cas} $ and obtained strong binding
due to large contribution of $\Delta M_1(p_1,q_1;Y_1,I_1)$ and $\Delta M_1(p_2,q_2;Y_2,I_2)$.
This very large binding seemed apparently unrealistic, and reasonable way out of this
situation appeared when it was recognized that the contributions of the order of $N_c^0$
due to poorly known loop corrections, or Casimir energy make large negative contribution
both to $M_1$ \cite{mous,mewa} and, probably, to $M_2$.
To obtain the $NN$ singlet scattering state on the right place, we should have \cite{kss2}
$$2M_1^{Cas}-M_2^{Cas} \simeq - 820\;MeV, \eqno (24)$$
for the choice of parameters made in \cite{kss2}, 
and results shown in Table 4 follow immediately.
Up to now these contrbutions to classical masses of skyrmions were calculated very approximately
only for the unit ($B=1$) skyrmion \cite{mous,mewa}. These contributions are negative
$M_1^{Cas} \sim -1\,GeV$, i.e. they act in right direction. For larger baryon numbers Casimir energy has not been 
calculated yet, because it is very nontrivial computational problem. 

Prediction of the $S=-3$ dibaryons with $(J^P;\,I)=(1^+,2^+;1/2)$ below the $\Lambda\Xi$
threshold has been made long ago by Goldman et al \cite{goldm} within a variant of the MIT bag 
model.
Recently strong attraction was found in some two-baryon channels with strangeness $S=-3$
and $-4$, in the leading order of chiral effective field theory, 
suggesting the possible existence of bound states \cite{haimei}.
Latest studies of strange dibaryons within quark models are presented in \cite{chpw} and
references therein.

To get spectrum of strange dibaryons in our chiral soliton approach we should transform 
basic formula $(16)$ for the quantum correction to the energy of multiskyrmions to 
$$\Delta M = |S|\omega_S+ {1 \over 2\Theta_I}\Bigl[c I_r(I_r+1)+(1-c)I(I+1) +(\bar c -c)I_S(I_S+1)\Bigr]+$$
$$ + {J(J+1)\over 2\Theta_J} +{\kappa^2\over 2}
\left({1\over \Theta_{I,3}}-  {1\over \Theta_{I}}- {4\over \Theta_{J}}\right), \eqno (25)$$
and $B=2$ in all quantities $\Theta,\, \omega_S$ to be taken from Table 1. 
$I_r$ (the right isospin within the rigid rotator quantization scheme) is the isospin of 
the nonstrange state, $I_S\leq |S|/2$ is the isospin carried by strange mesons,
and the observed isospin $\vec I = \vec I_r + \vec I_S$.
For $S=0$ and $I=I_r$ we recover the above formula $(9)$ for the quantum correction 
to the $SU(2)$ quantized dibaryons.}\\

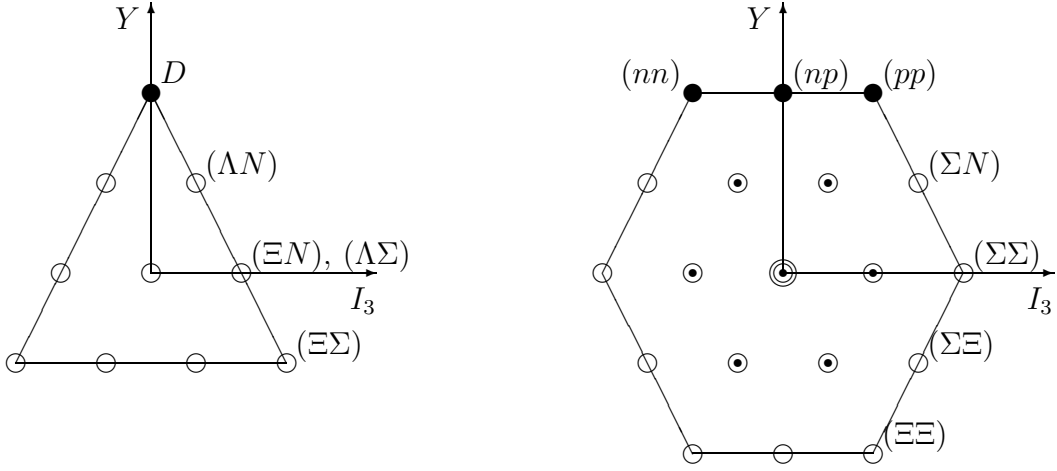
\begin{figure}[h]
\label{multiplet}
\setlength{\unitlength}{1.2cm}
\begin{flushleft}
\begin{picture}(12,5.5)
\put(3,3){\vector(1,0){2.5}}
\put(3,3){\vector(0,1){3}}
\put(2.6,5.7){$Y$}
\put(5.2,2.6){$I_3$}
\put(1,0){$B=2,\;\{\overline {10}\},\, J=1 \,(2,...)$}
\put(3.1,5.1){$ D$}
\put(3.6,4.1){$ (\Lambda N)$}
\put(4.1,3.1){$ (\Xi N),\,(\Lambda\Sigma)$}
\put(4.6,2.1){$ (\Xi\Sigma)$}

\put(3,5){\circle*{0.21}}
\put(2.5,4){\circle {0.2}}
\put(3.5,4){\circle{0.2}}
\put(2,3){\circle{0.2}}
\put(3,3){\circle{0.2}}
\put(4,3){\circle{0.2}}
\put(1.5,2){\circle{0.2}}
\put(2.5,2){\circle{0.2}}
\put(3.5,2){\circle{0.2}}
\put(4.5,2){\circle{0.2}}

\put(1.5,2){\line(1,0){3}}
\put(1.5,2){\line(1,2){1.5}}
\put(4.5,2){\line(-1,2){1.5}}

%end of anti-10
\put(10,3){\vector(1,0){3}}
\put(10,3){\vector(0,1){3}}
\put(9.6,5.7){$Y$}
\put(12.7,2.6){$I_3$}
\put(8,0){$B=2,\;\{27\},\, J=0\,(1,...)$}
\put(8.2,5.1){$ (nn)$}
\put(10.1,5.1){$(np)$}

\put(11.1,5.1){$(pp) $}
\put(11.6,4.1){$(\Sigma N) $}
\put(12.1,3.1){$(\Sigma\Sigma) $}
\put(11.6,2.1){$(\Sigma\Xi) $}
\put(11.1,1.1){$(\Xi\Xi) $}

\put(9,5){\circle*{0.21}}
\put(10,5){\circle*{0.21}}
\put(11,5){\circle*{0.21}}

\put(8.5,4){\circle{0.2}}
\put(9.5,4){\circle*{0.1}}
\put(9.5,4){\circle {0.2}}
\put(10.5,4){\circle*{0.1}}
\put(10.5,4){\circle {0.2}}
\put(11.5,4){\circle{0.2}}

\put(8,3){\circle{0.2}}
\put(9,3){\circle*{0.1}}
\put(10,3){\circle*{0.1}}
\put(11,3){\circle*{0.1}}
\put(12,3){\circle{0.2}}
\put(9,3){\circle {0.2}}
\put(10,3){\circle {0.2}}
\put(11,3){\circle {0.2}}
\put(10,3){\circle {0.3}}

\put(8.5,2){\circle{0.2}}
\put(9.5,2){\circle*{0.1}}
\put(9.5,2){\circle {0.2}}
\put(10.5,2){\circle*{0.1}}
\put(10.5,2){\circle {0.2}}
\put(11.5,2){\circle{0.2}}

\put(9,1){\circle{0.2}}
\put(10,1){\circle{0.2}}
\put(11,1){\circle{0.2}}

\put(8,3){\line(1,2){1}}
\put(8,3){\line(1,-2){1}}
\put(9,5){\line(1,0){2}}
\put(9,1){\line(1,0){2}}
\put(12,3){\line(-1,2){1}}
\put(12,3){\line(-1,-2){1}}

%end of 27-plet
\end{picture}
\end{flushleft}\end{figure}
{\tenrm{\bf Figure 1:} $I_3-Y$ diagrams of multiplets of dibaryons $B=2$. 
Virtual levels (scattering states) are shown in brackets, e.g. $(\Lambda N)$ scattering 
state which appears as near threshold enhancement.}\\

For the difference of energies of states which belong to antidecuplet and singlet $(NN) \; ^1S_0$ 
state we obtain
$$ E (0,3; I,J,S)- E(2N, ^1S_0) = |S|\omega_S + {\mu_{S,B}-1 \over 4\mu_{S,B} \Theta_{S,B}}
I(I+1)+$$
$$ + {(\mu_{S,B}-1)( \mu_{S,B}-2) \over 4\mu_{S,B}^2 \Theta_{S,B}}I_S(I_S+1)
+{1\over 2\Theta_J}J(J+1)-{1\over \Theta_I}+\kappa^2 \delta(\Theta), \eqno (26)$$
and in our case of $S=-1$ we should take $I_S=1/2$. The only allowed possibility for 
$\kappa$ is $\kappa=0$, because $I_r=0$. Numerical values of dibaryons energies 
are given for several lowest states in Table 5.

\begin{center}
\begin{tabular}{|l|l|l|l|l|l|l|}
\hline
$ I_r$ &$\, J$ & $\;I$ &$\; S $& $ \, \kappa $& $\Delta E(MeV) $\\
\hline
$ 0$& $\, 1$ &$ 1/2$&$\; -1 $&$\; 0$ &$\;\; 289  $ \\
\hline
$ 0$& $\, 2$ &$ 1/2$&$\; -1 $&$\; 0$ &$ \;\;392  $ \\
\hline
$ 0$& $\, 3$ &$ 1/2$&$\; -1 $&$\; 0$ &$ \;\;546  $ \\
\hline

\end{tabular}
\end{center}

{\tenrm{\bf Table 5.} $B=2$ states: set of quantum numbers and the energy above the $NN$
scattering state for the $S=-1$ states with $I_r=0$ and different values of spin, 
to be ascribed to antidecuplet, $(p,q)=(0,3)$, shown in Fig.1a.}\\

The state with $J=I_r=0$, $S=-1,\;I=1/2$, not shown in Table 5, has energy $\Delta E(0,0,1/2,-1) \simeq 238\,MeV$.
but this state cannot belong to the antidecuplet containing deuteron with $J=1$.

For dibaryon states which belong to $\{27\}$-plet we can use Eq. $(20)$ with $I_r=1$,
$I_S=1/2$, $J_0=0$, $\kappa_0=0$. Numerical results are shown in Table 6.
%In our approach theoretical mass difference of $\Lambda$-hyperon and nucleon is
%$M_\Lambda - M_N \simeq 275\, MeV$ - too large in comparison with experimental value
%of $177\,MeV$.

We would like to stress again that we are not fitting --- here and previously --- 
the absolute values of masses of nucleons, hyperons and nuclei (in difference from papers \cite{anw,mmw})
because they are controlled by poorly known loop corrections or Casimir energy
(see discussion of Eq. $(24)$).

\begin{center}
\begin{tabular}{|l|l|l|l|l|l|l|}
\hline
$ I_r$ &$\, J$ & $\;I$ &$\; S $& $ \, \kappa $& $\Delta E(MeV) $\\
\hline
$ 1$&   $\, 0$ &$\, 1/2$&$\, -1 $&$\; 0$ &$\;\;262 $ \\
\hline
$ 1$&$\, 1$ &$\, 1/2$&$\, -1 $&$\; 0$ &$\;\;313 $ \\
\hline
$ 1$&$\, 1$ &$\, 3/2$&$\, -1 $&$\; 0$ &$\;\;357 $ \\
\hline
$ 1$&$\, 2$ &$\, 1/2$&$\, -1 $&$\; 0$ &$\;\;416 $ \\
\hline
$ 1$& $\, 2$ &$\, 1/2$&$\; -1 $&$\; 1$ &$\;\;339 $ \\
\hline
$ 1$& $\, 3$ &$\, 1/2$&$\; -1 $&$\; 1$ &$\;\;493 $ \\
\hline
\end{tabular}
\end{center}

{\tenrm{\bf Table 6.} $B=2$-states: set of quantum numbers and the energy above the $NN$
threshold for the $S=-1$ states with $I_r=1$, which can be ascribed 
to the $27$-plet, $(p,q)=(2,2)$, see Fig.1b.}\\

As we can see from Table 6, the state with isospin $I=3/2$ has greater mass than the state with $I=1/2$
and same other quantum numbers: $(J=1,\; S=-1,\;P=+1)$. The state with negative parity has smaller mass
than the state with positive parity, $J=2$.
\begin{center}
{\bf
\begin{figure}[h]
\label{B number 2}
\setlength{\unitlength}{1.2cm}
\begin{flushleft}
\begin{picture}(13,6.5)

\put(0.5,0){\line(0,1){7.}}
\put(0.5,1){\line(1,0){0.15}}
\put(0.5,2){\line(1,0){0.15}}
\put(0.5,3){\line(1,0){0.15}}
\put(0.5,4){\line(1,0){0.15}}
\put(0.5,5){\line(1,0){0.15}}
\put(0.5,6){\line(1,0){0.15}}
\put(0.5,8){\line(1,0){0.15}}
\put(-.5,0.8){$ 100$}
\put(-.5,4.8){$500$}

\put(6,5.1){$\bar K NN$}
\put(2.3,4.98){\line(1,0){5.}}
\put(2.3,4.965){\line(1,0){5.}}
\put(2.3,4.95){\line(1,0){5.}}

\put(6.5,-.5){$NN-threshold$}
\put(2.3,0.){\line(1,0){12.}}
\put(2.3,-0.015){\line(1,0){12.}}
\put(2.3,-0.03){\line(1,0){12.}}

\put(14.3,0){\line(0,1){1.}}
\put(14.5,0.7){$100$}
\put(14.5,0.2){$MeV$}

\put(2.5,4.93){\line(1,0){1.}}
\put(2.5,4.92){\line(1,0){1.}}
\put(2.5,4.91){\line(1,0){1.}}

\put(2.5,4.35){\line(1,0){1.}}
\put(2.5,4.33){\line(1,0){1.}}
\put(2.5,4.34){\line(1,0){1.}}

\put(2.5,3.39){\line(1,0){1}}
\put(2.5,3.38){\line(1,0){1}}
\put(2.5,3.37){\line(1,0){1}}

%\put(2,2.262){\line(1,0){1}}
\put(0.6,5.6){ $J,\;\,I,\;\; S$}
\put(0.6,4.73){ $3,1/2,-1$}
\put(0.6,4.71){ $3,1/2,-1$}

\put(0.6,4.08){ $2,3/2,-1$}
\put(0.6,4.06){ $2,3/2,-1$}

\put(0.6,3.24){ $2,1/2,-1$}
\put(0.6,3.26){ $2,1/2,-1$}

\put(2.3,0.5){$P=-1 $}
\put(2.3,0.1){$\kappa=1 $}
%end of P=-1

\put(6.2,4.14){\line(1,0){1.}}
\put(6.2,4.15){\line(1,0){1.}}
\put(6.2,4.16){\line(1,0){1.}}

\put(6.2,3.92){\line(1,0){1}}
\put(6.2,3.93){\line(1,0){1}}

\put(6.2,3.58){\line(1,0){1}}
\put(6.2,3.57){\line(1,0){1}}

\put(6.2,2.62){\line(1,0){1}}
\put(6.2,2.63){\line(1,0){1}}

\put(6.2,2.89){\line(1,0){1}}
\put(6.2,2.90){\line(1,0){1}}

\put(4.2,4.2){ $2,1/2,-1$}
\put(4.2,3.74){ $2,1/2,-1$}

\put(4.2,3.34){ $1,3/2,-1$}

\put(4.2,2.8){ $1,1/2,-1$}
\put(4.2,2.4){ $0,1/2,-1$}

\put(6.1, 0.5){$P=+1 $}
\put(6.1, 0.1){$\kappa=0 $}
%end of P=+

\put(9.4,6.06){\line(1,0){1.}}
\put(9.4,4.02){\line(1,0){1.}}
\put(9.4,2.49){\line(1,0){1.}}
\put(9.4,2.29){\line(1,0){1.}}
\put(9.4,0.75){\line(1,0){1.}}

\put(8.,5.76){ $4,2,0$}
\put(8.,3.72){ $3,2,0$}
\put(8.,2.6){ $2,2,0$}
\put(8.,1.98){ $3,1,0$}
\put(8.,0.62){ $2,1,0$}

\put(9.4,0.14){$P=-1 $}
%\put(9.4,0.02){$\, \kappa=1. $}
%end of P=-1, S=0

\put(12.8,4.62){\line(1,0){1.}}
\put(12.8,4.35){\line(1,0){1.}}
\put(12.8,2.25){\line(1,0){1.}}
\put(12.8,2.19){\line(1,0){1.}}
\put(12.8,1.53){\line(1,0){1.}}
\put(12.8,-0.36){\line(1,0){1.}}
\put(12.8,-0.34){\line(1,0){1.}}

\put(2.25, 1.74){- - - - - - - - - - - - - - - - - - -}
\put(2.45, 1.85){$\Lambda N (experim.)$}

\put(11.4,4.7){ $4,2,0$}
\put(11.4,4.25){ $0,3,0$}
\put(11.4,2.35){ $1,2,0$}
\put(11.4,1.86){ $3,0,0$}
\put(11.4,1.33){ $2,1,0$}
\put(11.4,-0.46){ $1,0,0$}
\put(11.4,0.04){ $0,1,0$}

\put(12.6, 0.14){$P=+1 $}
%\put(12.6, 0.02){$\,\kappa=0 $}
\end{picture}
\end{flushleft}
\vspace{0.2cm}
%\protect\caption{Position }
\end{figure}}
\end{center}

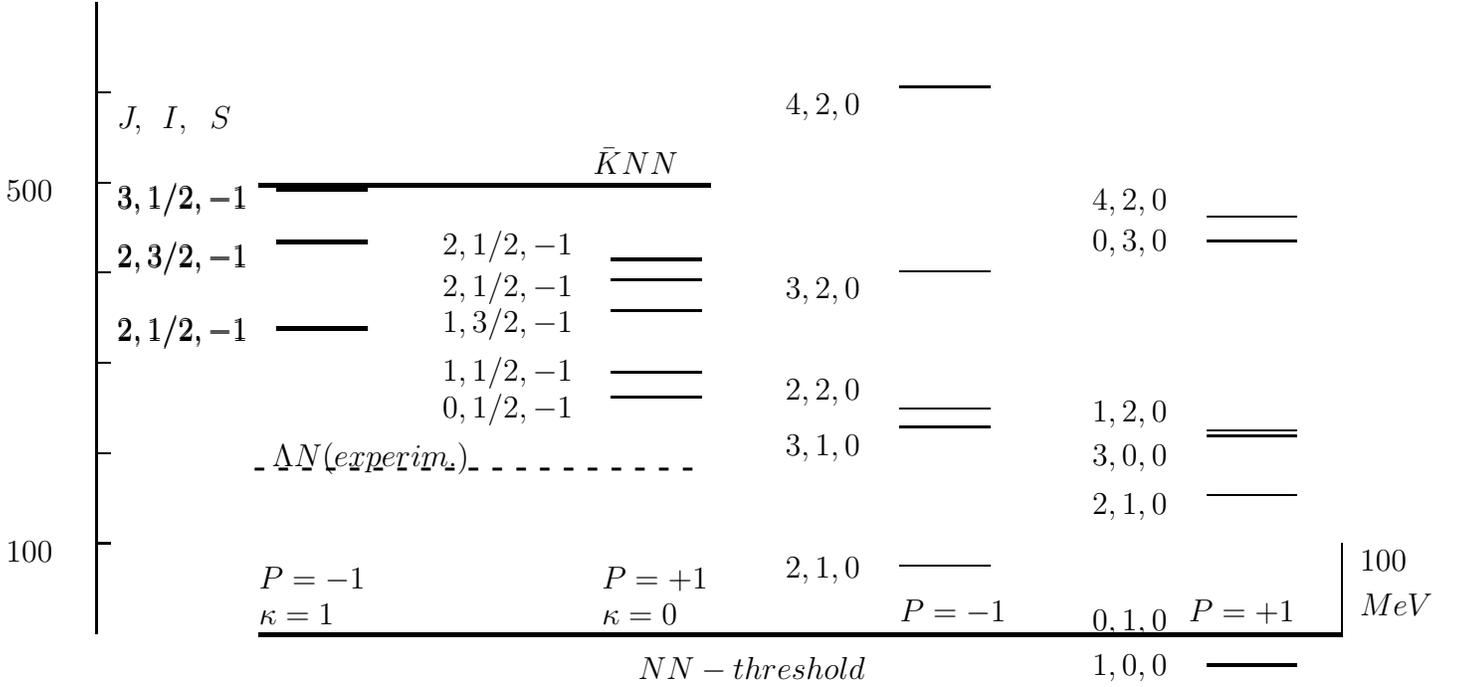
{\tenrm{\bf Figure 2:} Position of the $B=2$ states above the $NN$ threshold with negative 
strangeness, negative and positive parities (first 2 columns); with zero strangeness,
negative and positive parities (columns 3 and 4). 
The $\bar K NN$ threshold is shown by black line, as well as the $NN$ threshold.
The dashed line indicates the $\Lambda N$ threshold with empirical value of $M_\Lambda$.
The accuracy of calculation is not better than $\sim 40\,MeV$.}
\section{Some of the $B=3$, $S=-1$ states}
For the $B=3$ system the expression for the difference of energies (masses) of state 
with strangeness $S$, isospin $I$, spin $J$ and the ground state with zero strangeness, isospin $I_r$
is similar to $(23)$
$$\Delta E (p,q; I,J,S; I_r,J_0,0) \simeq |S|\omega_S + {\mu_{S,B}-1 \over 4\mu_{S,B} \Theta_{S,B}}
[I(I+1)-I_r(I_r+1)]+$$
$$ + {(\mu_{S,B}-1)( \mu_{S,B}-2) \over 4\mu_{S,B}^2 \Theta_{S,B}}I_S(I_S+1)
+{1\over 2\Theta'_J}\left[J(J+1)-J_0(J_0+1) + M^2{\Theta_{int}\over \Theta_I-\Theta_{int}}\right], \eqno (27)$$
with $\Theta_J' = (\Theta_J\Theta_I-\Theta_{int}^2)/(\Theta_I-\Theta_{int})$,
all quantities should be taken from Table 1 for $B=3$, $\Theta_{int}\simeq -9.4\, Gev^{-1}$.

For the ground $B=3$ state the $SU(3)$ multiplet with $(p,q)=(1,4),\; I_r=J_0=1/2$
($\overline{\{35\}}$-plet) is shown in Fig.3a. Fig.3b for even $B$-numbers is included for illustration.
The equality $J_0=I_r$ follows from the symmetry properties of the $B=3$ classical configuration
which has tetrahedral form. see \cite{c3}. 

\def\br{\mbox{\boldmath $r$}}
\def\bm{\mbox{\boldmath $m$}}
\setlength{\unitlength}{1.25cm}
\begin{flushleft}

\begin{picture}(12,6.)
\put(3,.5){\vector(1,0){2.5}}
\put(3,.5){\vector(0,1){3.8}}
\put(2.65,4.2){ $Y$}
\put(5.5,0.2){\bf $I_3$}
\put(1.,4.7){ a) $Odd \;B\,,\; J=1/2$}
\put(2.3,3.2){$^3 H$}
\put(3.3,3.2){$^3 He$}
\put(2.8,2.3){$^3_\Lambda H$}
%\put(3,4){\circle*{0.1}}
\put(2.5,3){\circle* {0.15}}
\put(3.5,3){\circle* {0.15}}
\put(2,2){\circle {0.15}}
\put(3,2){\circle* {0.15}}
\put(3,2){\circle {0.27}}
\put(4,2){\circle {0.15}}
\put(1.5,1){\circle {0.15}}
\put(2.5,1){\circle {0.15}}
\put(3.5,1){\circle {0.15}}
\put(4.5,1){\circle {0.15}}

%\put(1.5,1){\line(1,0){3}}
%\put(1.5,1){\line(1,2){1.5}}

%\end{picture}
%\begin{picture}(6,6)

\put(9,.5){\vector(1,0){2.3}}
\put(9,.5){\vector(0,1){3.8}}
\put(8.8,4.2){$Y$}
\put(11,0.2){$I_3$}
\put(7.,4.7){ b) $Even \;B\,,\; J=0$}
\put(8.8,3.2){$^4 He $}
\put(8.3,2.3){$^4_\Lambda H$}
\put(9.3,2.3){$^4_\Lambda He$}

\put(9,3){\circle* {0.15}}
%\put(9.5,3){\circle {0.2}}

%\put(7.5,1){\circle*{0.1}}
\put(8.5,2){\circle*{0.15}}
\put(8.5,2){\circle {0.27}}
\put(9.5,2){\circle*{0.15}}
\put(9.5,2){\circle {0.27}}
%\put(10.5,1){\circle {0.15}}

\put(8,1){\circle {0.15}}
\put(9,1){\circle {0.15}}
\put(10,1){\circle {0.15}}
\end{picture}
\end{flushleft}
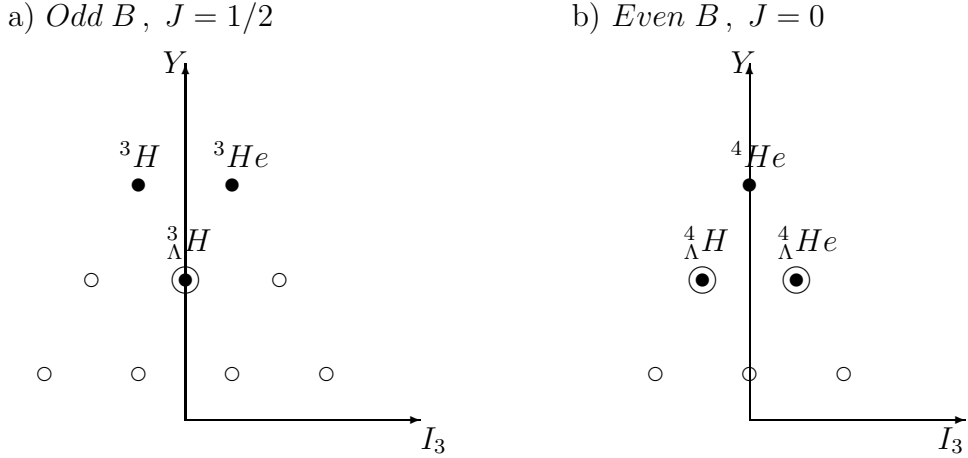
{\tenrm{\bf Figure 3:}  (a) The location of the isoscalar ground state (shown by double circle)
with odd $B$-number and $S=-1$ in the upper part of the $(I_3 -Y)$ diagram.
(b) The same for isodoublet states with even $B$. The case of light hypernuclei
$_\Lambda H$ and $_\Lambda He$ is presented as an example. The lower parts of 
diagrams with $Y \leq B-3$ are not shown here.}\\

\begin{center}\begin{tabular}{|l|l|l|l|l|l|l|}
\hline$ I_r$ &$\;J  $ &$\,I$ & $ \, M $& $\Delta E(MeV) $\\
\hline$ 1/2$&$\, 1/2$ &$\, 0$&$\; 0$ &$\;279 $ \\
\hline$ 1/2$&$\, 1/2$ &$\, 1$&$\; 0$ &$\;321 $ \\
\hline$ 3/2$&$\, 3/2$ &$\, 1$&$\; 0$ &$\;378 $ \\
\hline$ 3/2$&$\, 3/2$ &$\, 2$&$\; 0$ &$\;462 $ \\
\hline
\hline$ 3/2$&$\, 3/2$ &$\, 1$&$\; 3$ &$\; 302$ \\
\hline$ 3/2$&$\, 3/2$ &$\, 1$&$\; 2$ &$\; 348$ \\
\hline$ 3/2$&$\, 3/2$ &$\, 2$&$\; 2$ &$\; 432$ \\
\hline$ 5/2$&$\, 5/2$ &$\, 2$&$\; 4$ &$\; 421$ \\
\hline$ 5/2$&$\, 5/2$ &$\, 2$&$\; 3$ &$\; 482$ \\
\hline\end{tabular}\end{center}
{\tenrm{\bf Table 7.} Some of possible $B=3$-states: set of quantum numbers and the energy above the 
$NNN$ threshold for states with $I_r=1/2$, $3/2$ and $5/2$, strangeness $S=-1$ and different values of spin, 
isospin, and parity, $M=M_3$.}\\

Our results for $S=-1$ excited tribaryons are presented in Table 7. 
The lowest in energy state with $J=I_r=1/2$, $I=M=0$ can be naturally interpreted
as $^3_\Lambda H$ hypernucleus. States with $J=3/2$ and $5/2$ should belong to other $SU(3)$ multiplets.

These results should be
considered as preliminary; further studies of this issue are desirable, also
for greater baryon numbers, $B\geq 4$.

\newpage
\begin{center}{\bf\begin{figure}[h]\label{B number 3}
\setlength{\unitlength}{1.15cm}\begin{flushleft}\begin{picture}(13,6.5)
\put(0.5,0){\line(0,1){7.}}
\put(0.5,1){\line(1,0){0.15}}
\put(0.5,2){\line(1,0){0.15}}
\put(0.5,3){\line(1,0){0.15}}
\put(0.5,4){\line(1,0){0.15}}
\put(0.5,5){\line(1,0){0.15}}
\put(0.5,6){\line(1,0){0.15}}
\put(0.5,8){\line(1,0){0.15}}
\put(-.5,0.8){$100$}
\put(-.5,4.8){$500$}
\put(6,5.1){$\bar K NNN$}
\put(3.8,4.98){\line(1,0){6.5}}
\put(3.8,4.965){\line(1,0){6.5}}
\put(3.8,4.95){\line(1,0){6.5}}

\put(3.8,1.76){- - - - - - - - - - - - - - - - - - - - - - - - -}

\put(5.,1.91){$\Lambda NN (experim.)$}

\put(6.5,-.5){$NNN-threshold$}
\put(3.8,0.){\line(1,0){9.}}
\put(3.8,-0.015){\line(1,0){9.}}
\put(3.8,-0.03){\line(1,0){9.}}
\put(12.8,0){\line(0,1){1.}}
\put(12.,0.7){$100$}
\put(12.,0.2){$MeV$}

\put(4,3.02){\line(1,0){1.}}
\put(4,3.03){\line(1,0){1.}}
\put(4,3.04){\line(1,0){1.}}

\put(4,3.48){\line(1,0){1.}}
\put(4,3.50){\line(1,0){1.}}
\put(4,3.51){\line(1,0){1.}}

\put(4,4.32){\line(1,0){1.}}
\put(4,4.33){\line(1,0){1.}}
\put(4,4.33){\line(1,0){1.}}

\put(4,4.21){\line(1,0){1.}}
\put(4,4.22){\line(1,0){1.}}
\put(4,4.23){\line(1,0){1.}}

\put(4,4.82){\line(1,0){1.}}
\put(4,4.83){\line(1,0){1.}}
\put(4,4.84){\line(1,0){1.}}

%\put(4,3.48){\line(1,0){1.}}
%\put(4,3.50){\line(1,0){1.}}
%\put(4,3.51){\line(1,0){1.}}

%\put(2,2.262){\line(1,0){1}}
\put(1.3,5.6){ $J,\;\;\;I,\;\;\;M$}

\put(1.3,2.6){ $3/2,\,1,\;\;\;3$}

\put(1.3,3.3){ $3/2,\,1,\;\;\;2$}

\put(1.3,3.8){ $5/2,\,2,\;\;\;4$}

\put(1.3,4.2){ $3/2,\,2,\;\;\;2$}

\put(1.3,4.75){ $5/2,\,2,\;\;\;3$}

\put(3.8,0.5){$P=-1$}
%\put(2.3,0.1){$M_3=2$}
%end of P=-1
\put(9,2.79){\line(1,0){1.}}

\put(9,3.21){\line(1,0){1}}

\put(9,3.78){\line(1,0){1}}

\put(9,4.62){\line(1,0){1}}

\put(6.9,2.67){$1/2,\;0,\;0$}
\put(6.9,3.12){$1/2,\;1,\;0$}
\put(6.9,3.69){$3/2,\;1,\;0$}
\put(6.9,4.53){$3/2,\;2,\;0$}

\put(8.8, 0.5){$P=+1 $}
\put(8.8, 0.1){$M=0 $}
%end of P=+
%\put(11.4,6.06){\line(1,0){1.}}
%\put(10.,5.76){ $4,2,0$}
%\put(11.4,0.14){$P=-1 $}
%end of P=-1, S=0
%\put(12.8,4.35){\line(1,0){1.}}
%\put(12.8,3.73){\line(1,0){1.}}
%\put(12.8,2.16){\line(1,0){1.}}
%\put(12.8,1.53){\line(1,0){1.}}
%\put(12.8,-0.36){\line(1,0){1.}}
%\put(12.8,-0.34){\line(1,0){1.}}
%\put(11.4,4.15){ $0,3,0$}
%\put(11.4,3.53){ $4,2,0$}
%\put(11.4,1.96){ $3,0,0$}
%\put(11.4,1.33){ $2,1,0$}
%\put(11.4,-0.46){$1,0,0$}
%\put(12.6, 0.14){$P=+1 $}
%\put(12.6, 0.02){$\,\kappa=0 $}
\end{picture}
\end{flushleft}\vspace{0.5cm}
%\protect\caption{Position}}
\end{figure}}
\end{center}
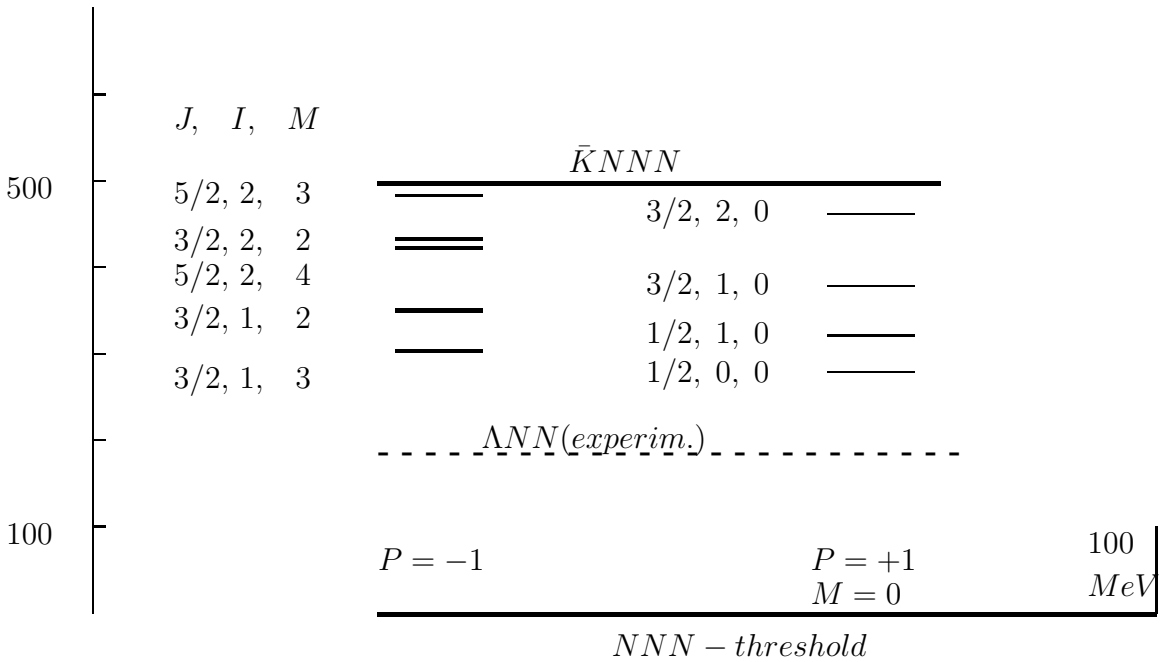
{\tenrm{\bf Figure 4:} Some of the $B=3$ rotationally excited states (above the $NNN$ threshold) with negative 
strangeness $S=-1$, different isospin and spin, negative and positive parities, $I_r=J$.}\\

The restriction on allowed isospin of non-exotic states (i.e. the states without additional quark-antiquark pairs) 
takes place: $I\leq  (3B+S)/2$, and for antikaon-nuclei bound states, evidently, $I\leq (B+1)/2$.
Second restriction becomes stronger for $B\geq 2$, so, only states with not too large isospin 
can be interpreted as antikaon-nuclei bound states.
Generally, rotational excitations have additional energy $\Delta E = {J(J+1)/2 \Theta_J}. $
The orbital inertia grows fast with increasing baryon (atomic) number,$\Theta_J \sim B^p$,
$p$ is between 1 and 2.   
By this reason the number of rotational states becomes large for large baryon numbers.
\section{Summary and conclusions}
To summarize, we have considered here rotational-type excitations of the $S=-1$ baryonic
systems (nuclei) with baryon number $B=2$ and, partly, $B=3$ using the chiral soliton
approach.
It was assumed that during the collective motion the shape of the basic classical 
configuration is not changed. We did not consider the vibration-breathing 
excitations which are possible as well. For the baryon 
number $1$ it was possible to describe in this way some properties of the negative parity $\Lambda (1405)$-state  \cite{scscg}.
For the case of dibaryons some nonstrange states have been considered in \cite{scsc}, although
numerical results have not been presented. Similar states should exist for strange multibaryons,
but numerical computations are extremely complicated.

We investigated only one of several possible variants of multiskyrmions quantization in the $SU(3)$ extension
of the chiral soliton model, the rigid rotator/oscillator variant.
The rich spectrum of strange multibaryons is predicted within this 
 approach, with positive as well as with negative parities.
There is rigorous theoretical statement that at small value of kaon mass there should be quantized 
states with strangeness $-1$ with energy below the $NN...+\bar K$ threshold.

The existence of strange excited nuclear states 
which could be interpreted as bound antikaon-nuclear states
within the CSA seems to be quite natural and not unexpected.
However, when the energy below the threshold becomes large, interpretation of such states as the bound
state of antikaon and corresponding nonstrange nucleus becomes less straightforward, due to 
increase of the weight of other components, first of all, containing hyperons. 
For realistic value of the kaon mass some states are predicted, but with considerable 
numerical uncertainty.
In view of these uncertainties, experimental investigations
could play decisive role. Since several such states are expected in the energy gap 
equal to one kaon mass, good 
enough experimental resolution in their energy (mass) of the observed states is of great importance.
Another option can be that there are several wide overlapping states, and in this case
better resolution will not help much.

Results of this paper have been presented partly at the 10th International Conference on
Hypernuclear and Strange Particle Physics (Hyp X),
Sep. 14-18, 2009 (Tokai, Ibaraki, Japan),
 VK is indebted to E.Oset for useful discussion at this conference.
The work has been supported by Fondecyt (Chile), 
grant number 1090236.
\\

\elevenbf{References}
\vglue 0.3cm

\end{document}